# A Multiple Linear Regression Approach For Estimating the Market Value of Football Players in Forward Position

Yunus Koloğlu, Hasan Birinci, Sevde Ilgaz Kanalmaz, Burhan Özyılmaz

**Abstract**—In this paper, market values of the football players in the forward positions are estimated using multiple linear regression by including the physical and performance factors in 2017-2018 season. Players from 4 major leagues of Europe are examined, and by applying Breusch – Pagan test for homoscedasticity, a reasonable regression model within 0.10 significance level is built, and the most and the least affecting factors are explained in detail.

**Index Terms**— Market Value Estimation, Football Statistics, Multiple Linear Regression,

## 1- INTRODUCTION

Linear regression is a statistical analysis which depends on modeling a relationship between two kinds of variables, dependent(response) and independent(predictor). The main purpose of regression is to examine if the independent variables are successful in predicting the outcome variable and which independent variables are significant predictors of the outcome. In this study, a linear regression with multiple independent variables will be built, in order to seek relevant factors that affect the market value of a football player in forward position. For this study, footballers from Europe leagues will be examined within their physical properties and their performance between 2017 and 2018 season. Both league and cup performances of the footballers are taken, and the data is gathered from the Transfermarkt.Although the transfers are done in the middle and in the end of the season, market values of the football players are updated properly. The most recent market values up to June 2018 are taken into the account, and detailed data about performance from the footballers who had played a full season in their league in the age of between 20 and 33 are collected from the forward players. Limitations of this study is between 2017 and 2018 season, as several times in the history, individual transfers were able to change the market value of the players significantly.

## 2- LITERATURE REVIEW

For this study, articles about the statistical analyses in sports branches are examined within the different variations of multiple linear regression. Brown [1] conducted a study on 86 College Football Teams in order to investigate rent of the players. He claimed that a team's number of premium players contributes to the team's overall performance and they both effect the team revenues in which marginal revenue product of a player is highly dependent on. At premium player stage, number of National Football League(NFL) Draftees from each team is used as a dependent variable and in the next stage, team performance, these NFL Draftees are used as an independent variable in order to predict teams rating(performance). At last, both team performance and NFL Draftees are used as a dependent variable to predict the team's revenues, which in return will give the rent of each player accordingly.

Carling Opta Index is a performance-based monitoring system which records and scores over 300 different actions that are used in calculating final game score for a player. The average of these last 6 game scores are taken as final game score index in order to be used in the model. By using this index, Tunaru [2] did an investigation about an option pricing framework for valuation of football players. There is a differential equation on how to turn this index to cash. He also adds unexpected events such as injuries into the model as Poisson processes in order to investigate long-term value of a player. The value of a player also depends on club because of total performance points of entire team.

Selection of bias is an error type caused by that randomization is not achieved while the selection of individuals, groups or data. Therefore, the sample is not representative of the population which was going to analysed. In order to avoid this situation, James Heckman did a study and came up with Heckman Correction model. Carmichael [3] studied the price structure of transfer market for professional players in English football leagues. He also used Heckman two-step procedure to take account of selection bias. The findings of his study claim that the probability of transfer is highest for more experienced players who can score.

Fullard [4] searched for two questions in his paper. One of the questions was to check whether there is a relation between structure types of salary and predicted team performance in National Hockey League. Fullard used a method which is Ordinary Least Squares to decrease the summation of squares of vertical gaps of observed y-axis values from the regression line. In this paper, multiple linear regression was used and market value of forward football players was taken as dependent variable. Fullard focused the salary of hockey goalies which means goalkeeper in football terms. While searching the reason why the market value assigned to a player, it is important to check whether there is a correlation between market value and contribution to the score. Fullard checked the similar relation between goalie's salary and the number cleared the ball. However, the results were not that expected. There is no correlation between salary and shootouts, shootout wins or shootout losses.

While Fullard studies on salaries of specifically goalies who play in National Hockey League, Peck [5] focused on the salaries of National Hockey League for all positions of a team. While looking for the reasons why a player earns that much salary, or their market value is a certain number, there are important performance measurements which are the contribution to the score, assist numbers, age, strong foot, played club, sponsor, and ethnicity. Peck used some independent variables are team performance, the number of careers game played, the number of assists and the number of goals. There are astonishing results obtained by using ordinary least squares. As long as determining the salaries of players, it differs position by position and it is expected that the salary of forward players is much higher than defence ones. However, the results reveal that a defenseman can earn much more than his forward teammate. In this paper, the model was limited by the market value of forward players in football. Peck's paper, also supports that multiple linear regression is adequate to determine if there exists a correlation between the market value of a forward football player and contribution to the score, ethnicity or whatever determined at the beginning of this paper.

The thesis by Louivion and Pettersson [6] has proximity to our research with use of multiple linear regression to affect performance on players' salary. The objective of the project was to analyse and detect which performance measures affect the salary of the players by examining statistical data. It examines the factors of salary effect for NBA's basketball which includes performance measures and salary data. Study also includes 22 likely variables as performance measure. In this research it has salaries of NBA players' as dependent variable likely, for our research it is market value of football players in 4 different leagues. For this study it concluded with regression analysis that 12 different covariates that has effect on players' salaries and the explanatory level of its regression model is 57.4%. The observation of the covariates shows that the most effective variable o the salaries is attempt of 2 and 3-point shots.

The paper from Castillo Ramirez et al. [7] examines the influence of the single player in the match score by using multiple linear regression model. The reason behind of using the model is that it concentrates on explanation of the behaviour of many dependent variables from a series of independent variables connected through a linear equation with a series of coefficients adequate to the number of independent variables statistically significant. The model must be satisfying the stipulation that concerned by the analysis of the residues or errors of the model, predict errors of the model must pursue normal distribution with average equals to zero.

In his study, Newman [8] examines the market value factors of players that are transferred into English Premier League in the period of 2009-2015, in order to detect whether there was any superstar effect in these players' market values or not. Superstar effect refers to the phenomenon where the success of a famous athlete decreases the performance of other athletes in the same competition. To see the effect quantile regression used for seeing how a player who is in the top 5% change the value of an incremental goal of goals scored compared to a player who scores the median number of goals. Due to Newman's OLS regression model; goals, assist, age, and minutes played have positive coefficient and significant while due to the hit performance peak of the players in late 20's age square has negative coefficient. The study concludes with evidence of the superstar effect considering the goals scored by forward players which are elite players worth more to their market value than an average player.

According to Farrar [9], multicollinearity occurs when two independent variable effects each other. In other words, one independent variable is dependent on another independent variable, then correlation matrix of dependent variables approaches to singularity. When correlation coefficient of this pair is +1 or -1 it is called perfect multicollinearity, in this case, even the coefficient close to these numbers, one of the variables should be removed from the model. Kendall proposed the first method to avoid multicollinearity as 'Artificial Orthogonalization' and McCallum [10] contributed to this method.

While implementing multiple linear regression test to the different variable, there must exist some requirements such as fixed coefficients and homoscedastic disturbances. Breusch and A. R. Pagan [11] explained that when the requirements above are not satisfied the interpretation of multiple linear regression results may be not correct. Breusch and Pagan offers two least squares regressions instead of implementing likelihood ratio test as an ordinary condition. However, it can be calculated by two least squares regressions. Getting rid of use of repetitive computations are necessary to acquired maximum likelihood estimates of the constants in whole model is offered in the paper.

## 3- METHODOLOGY, FINDINGS AND RESULTS

For this study, data from the forward players in the age of between 20 and 34 are collected from Premier League, La Liga, Bundesliga and Serie A; containing the information about the league they are playing, the club they are playing, their ages, heights, dominant legs, nationalities, outfitter, matches played in that season, direct contribution to goal, card scores, and total time they have played in that season. Before applying the data preprocessing steps, necessary eliminations are done on the footballers; such as the footballers who changed their team in the mid-season are not included in this study, as well as footballers who have market value less than €20 million, and the players who had played under 1,000 minutes in the season. All the data is obtained from Transfermarkt in this study. Since most of the data is categorical, One-Hot Encoding from the Sci-Kit Learn library of Python is used to convert data into the unique matrices. Since there are around 100 footballers after the necessary operations, in order to not make huge attribute lists for each players, also some values are categorized. Instead of converting continuous variables into discrete variables, another possible approach would be scaling and using them, but it is illogical when real world is modeled. For instance; a young footballer lacks experience whereas an old footballer lacks performance, and a tall footballer lacks dribbling whereas a short footballer lacks physical dominance. For this logic, several variables are discretized. Although not doing it would make the model easier, it allows us to represent the idea of benefit of being in the middle ground.

Age is categorized in 7 basic groups, with starting from 20-21 and ending with over 31. Height is grouped in 7 categories, with starting from 160-164 and ending with over 190. Contribution to the goal is calculated as the sum of goals and half of the assists. Card scores are gathered in a single category with 1 point per yellow card, 2 points per 2 yellow cards in the same match, and 3 points for the red card. And finally, matches played in the season are grouped as increasing by 1 in every 5 matches, with 0 to 15 matches taking the value of 1. In terms of played matches, it can be assumed that there is a hidden lower limit of played matches in a season, as until a certain point, amount is not effective and there are several jump points, thus there is a monotonically increasing relation, hence the amount of matches are taken as discrete variables and encoded as well. According to built scale, player who had played 20 matches in a season gets the value of 2, 40 matches get the value of 6, 11 or 2 matches get the value of 1. Continuous variables in the model are standardized, whereas encoded variables are not.

In order to avoid dummy trap, 1 column from each attribute is deleted, and a bias unit consisting of ones is added in front of the data as $x_0$. In the initial model, there were 105 different attributes. As the model also included 105 samples, it was an unacceptable number, thus a feature selection with backward elimination method is used. In 0.1 significance level, there were 55 attributes and in 0.05 significance level, there were 25 different attributes. After that, in order to select the most appropriate significance level for the study, "Breusch-Pagan Test" is used to measure homoscedasticity / heteroscedasticity, which is the assumption of the level of all residuals in the model. As several data points in the model has higher residuals, the model becomes more heteroscedastic. Extraordinary points could pull the regressor from the center, as well as making the regressor biased and these are the main problems with heteroscedasticity. One way of reducing heteroscedasticity is using weighted least squares instead of OLS (Ordinary Least Squares), and other ways are applying Log Transformation and Box-Cox transformation. Since OLS in the MLR method in this study is used, we will try to find a combination of variables to have a model that fails to reject null hypothesis.

```
                            OLS Regression Results
==============================================================================
Dep. Variable:                      y   R-squared:                       0.963
Model:                            OLS   Adj. R-squared:                  0.575
Method:                 Least Squares   F-statistic:                     2.482
Date:                Mon, 04 Jun 2018   Prob (F-statistic):             0.0693
Time:                        08:26:11   Log-Likelihood:                -329.96
No. Observations:                 105   AIC:                             851.9
Df Residuals:                       9   BIC:                             1107.
Df Model:                          95
Covariance Type:            nonrobust
==============================================================================
                 coef    std err          t      P>|t|      [0.025      0.975]
------------------------------------------------------------------------------
const        -91.7510    114.958     -0.798      0.445    -351.803     168.301
x1           -13.4005     21.854     -0.613      0.555     -62.839      36.038
x2             5.8887     68.589      0.086      0.933    -149.272     161.049
x3             8.6394     14.820      0.583      0.574     -24.887      42.166
x4             6.8389     83.631      0.082      0.937    -182.347     196.024
x5            29.7442     43.778      0.679      0.514     -69.289     128.777
x6             0.5004     37.875      0.013      0.990     -85.178      86.179
x7             9.6291     55.354      0.174      0.866    -115.590     134.848
x8            41.6774     22.587      1.845      0.098      -9.418      92.773
x9           -21.5557     27.169     -0.793      0.448     -83.017      39.906
x10           32.3017     26.413      1.223      0.252     -27.449      92.052
x11           16.5779     36.353      0.456      0.659     -65.659      98.815
x12          -29.1851     32.016     -0.912      0.386    -101.611      43.241
x13            3.9024     36.357      0.107      0.917     -78.344      86.149
x14          -52.0604     52.662     -0.989      0.349    -171.190      67.069
x15           21.2306     73.567      0.289      0.779    -145.191     187.652
```

**Figure 1:** Part of regression summary of data without feature selection

When the model is run within 105 attributes, an $R^2$ value 0.963, and adjusted $R^2$ value of 0.575 is obtained, as shown in table. In addition to that, covariance type in OLS summary is classified as nonrobust, meaning that model is incapable of handling multicollinearity. MAPE error is 17%, and Breusch – Pagan test significance value is 0.897 for the model, which fails to reject the null hypothesis of homoscedasticity. Multicollinearity happens in the situations where there one of the independent variable can be guessed by the combination of several other variables in the model. It causes same regression line to have different equations, with different coefficients, thus it makes it harder to interpret the effect of independent variables in the model. Elimination by the variance inflation could be applied partially in the dataset, whereas we again used feature selection in that situation. VIF value tells us how much of the variance is inflated in each coefficient. As a rule of thumb, VIF value of 1 means it is not correlated, between 1 and 5 means it is moderately correlated, and more than 5 means highly correlated.

$$Variance\ Inflation\ Factor = \frac{1}{1 - R_i^2}$$

**Equation 1:** Formula of Variance Inflation Factor

$$MAPE = \frac{\sum_i^n |Y_i - \hat{Y}_i|}{Y_i}$$

**Equation 2:** Formula of Mean Absolute Percentage Error

Figure 2 and figure 3 shows the residual and actual/predicted plots for the data. Residual plot shows the homoscedasticity in the model, and actual/predicted plot show that although there are a little bit outliers, model shows a central tendency.

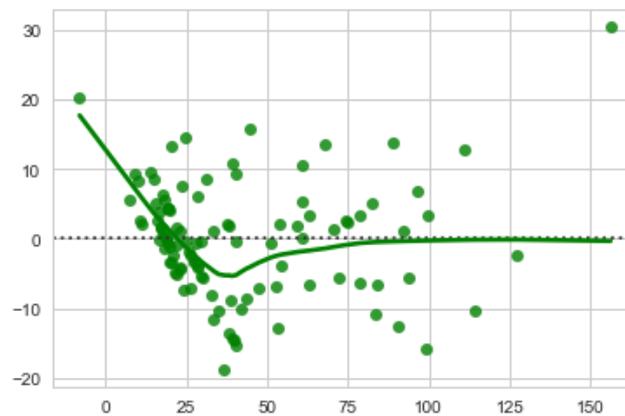

**Figure 2:** Residual plot of data without feature selection

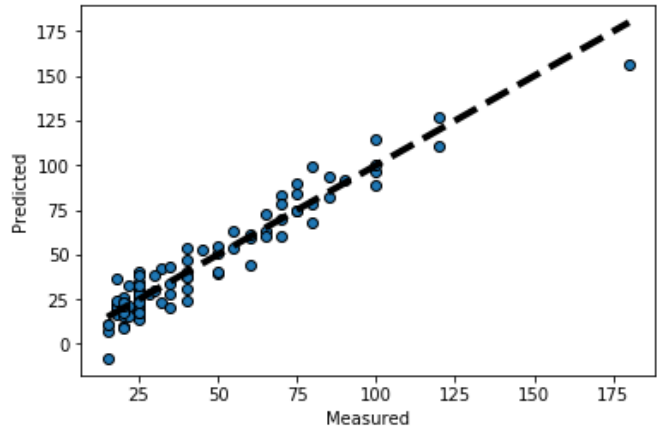

**Figure 3:** Measured / Predicted plot of data without feature selection

A backward elimination is applied in the data with both significance levels of 0.05 and 0.1 to compare the models. A simpler model is generally more prone to underfitting, less complex and time consuming, has less variance and more bias while a complex model is generally more prone to underfitting, more complex and time consuming and has a high variance but less bias. It is important to decide the best level of complexity in order to build a robust model.

As seen in figure 4, after the feature selection process, the amount of variables had reduced to 52, with $R^2$ value had decreased from 0.963 to 0.93 whereas adjusted $R^2$ had increased to 0.86 from 0.575. MAPE value is %20. It is also a good indicator to have a higher adjusted value as $R^2$ could be biased. Significance of BP had decreased to 0.44 from 0.897, but still it fails to reject null hypothesis of homoscedasticity. Figure 5 shows the actual / predicted plot, and figure 6 shows the residual plot.

```
                            OLS Regression Results
==============================================================================
Dep. Variable:                      y   R-squared:                       0.930
Model:                            OLS   Adj. R-squared:                  0.860
Method:                 Least Squares   F-statistic:                     13.29
Date:                Mon, 04 Jun 2018   Prob (F-statistic):           4.05e-17
Time:                        08:57:21   Log-Likelihood:                -363.75
No. Observations:                 105   AIC:                             833.5
Df Residuals:                      52   BIC:                             974.2
Df Model:                          52
Covariance Type:            nonrobust
==============================================================================
                 coef    std err          t      P>|t|      [0.025      0.975]
------------------------------------------------------------------------------
const         -49.4294     12.768     -3.871      0.000     -75.051     -23.808
x1             13.3396      7.285      1.831      0.073      -1.279      27.958
x2             35.7294     14.381      2.484      0.016       6.871      64.587
x3             51.3703      6.942      7.400      0.000      37.440      65.301
x4            -23.0743      7.936     -2.907      0.005     -39.000      -7.149
x5             23.2793      7.642      3.046      0.004       7.944      38.615
x6            -25.6298      8.807     -2.910      0.005     -43.303      -7.957
x7            -20.7825      8.644     -2.404      0.020     -38.129      -3.436
x8             27.0538      7.027      3.850      0.000      12.952      41.155
x9             31.7221     14.024      2.262      0.028       3.580      59.864
x10           -18.6637      9.786     -1.907      0.062     -38.301       0.973
x11            22.4513      6.597      3.403      0.001       9.213      35.690
x12           -27.5489     12.642     -2.179      0.034     -52.916      -2.182
x13            19.2354      6.954      2.766      0.008       5.281      33.190
x14           -62.7216     16.277     -3.853      0.000     -95.384     -30.059
x15            72.6052     13.081      5.550      0.000      46.355      98.855
x16            38.4339      9.173      4.190      0.000      20.028      56.840
```

**Figure 4:** Part of regression summary of the data within the feature selection at 0.1 significance

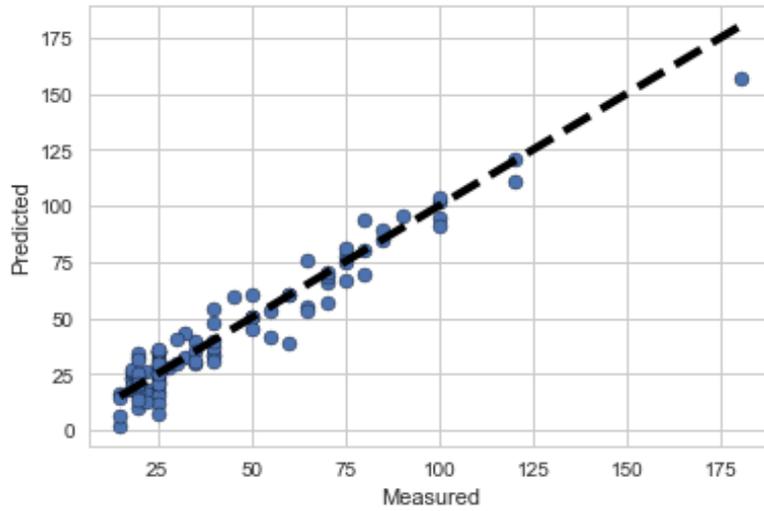

**Figure 5:** Measured / Predicted of the data within the feature selection at 0.1 significance

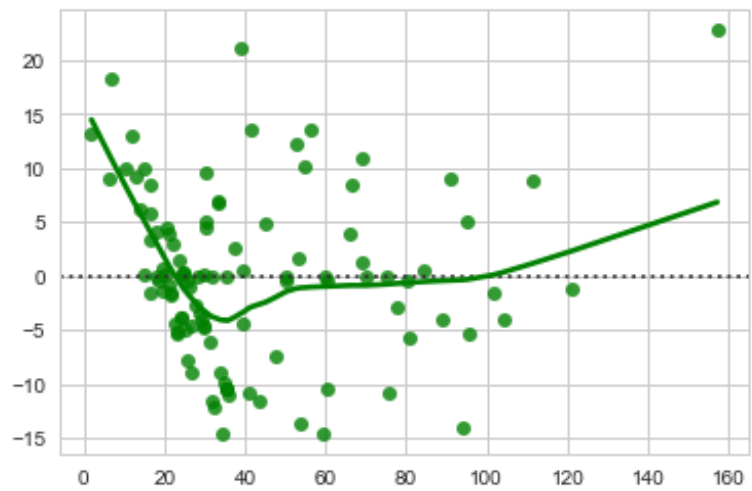

**Figure 6:** Residual plot of the data within the feature selection at 0.1 significance

When the significance level in Backward Elimination had decreased to 0.05, $R^2$ value had increased to 0.944 and adjusted $R^2$ value had increased to 0.927, as number of variables had decreased to 25. MAPE of the model is %27, but p value of the BP test had decreased to 0.003 with rejecting the null hypothesis of homoscedasticity. Figure 7 shows the regression results, figure 8 shows the measured / predicted plot and figure 9 shows the residual plot of the data. By graphs, homoscedasticity can't be observed easily, thus test value gains importance at that point.

```
==============================================================================
Dep. Variable:                      y   R-squared:                       0.944
Model:                            OLS   Adj. R-squared:                  0.927
Method:                 Least Squares   F-statistic:                     53.99
Date:                Mon, 04 Jun 2018   Prob (F-statistic):           4.30e-40
Time:                        09:14:28   Log-Likelihood:                -410.63
No. Observations:                 105   AIC:                             871.3
Df Residuals:                      80   BIC:                             937.6
Df Model:                          25
Covariance Type:            nonrobust
==============================================================================
                 coef    std err          t      P>|t|      [0.025      0.975]
------------------------------------------------------------------------------
x1            57.4574      7.957      7.221      0.000      41.622      73.293
x2           -22.3508     10.342     -2.161      0.034     -42.931      -1.770
x3            19.9860      7.284      2.744      0.007       5.490      34.482
x4            33.8492      6.505      5.204      0.000      20.905      46.794
x5           -28.8772     11.582     -2.493      0.015     -51.925      -5.829
x6            24.0523      6.619      3.634      0.000      10.879      37.225
x7           -34.7427     15.988     -2.173      0.033     -66.559      -2.927
x8            68.4578     14.596      4.690      0.000      39.412      97.504
x9            20.6690      7.771      2.660      0.009       5.204      36.134
x10           24.9943      8.345      2.995      0.004       8.387      41.601
x11           12.2860      5.860      2.097      0.039       0.625      23.948
x12           15.2947      4.639      3.297      0.001       6.062      24.527
x13           64.5626     21.038      3.069      0.003      22.696     106.429
x14           11.8840      4.083      2.911      0.005       3.759      20.009
x15           12.2368      3.411      3.587      0.001       5.449      19.025
x16           15.4844      7.583      2.042      0.044       0.395      30.574
x17           44.3643     18.197      2.438      0.017       8.151      80.577
x18           78.8110     14.831      5.314      0.000      49.296     108.326
```

**Figure 7:** Part of regression summary of the data within the feature selection at 0.05 significance

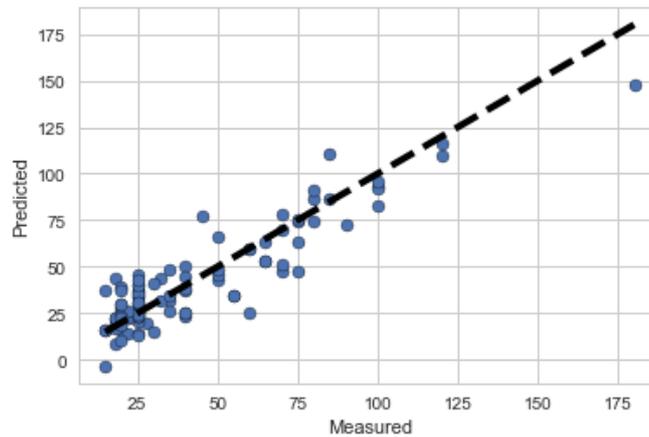

**Figure 8:** Measured / Predicted plot of the data within the feature selection at 0.05 significance

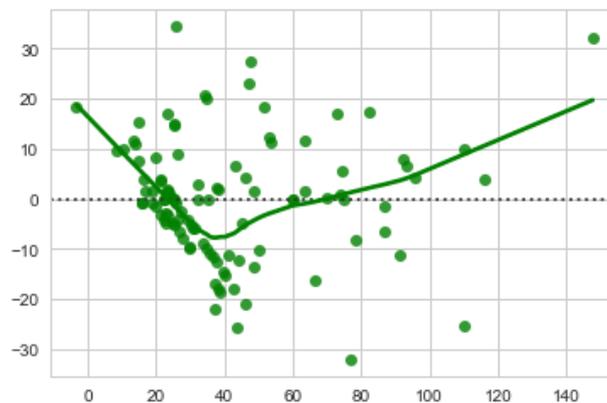

**Figure 9:** Residual plot of the data witnh the feature selection at 0.05 significance

With considering these criteria, the most appropriate model is the feature selected model with 0.1 significance level, with 52 attributes. However, still the collinearity between the variables couldn't be eliminated even with the minimal number of attributes. Among 52 affecting attributes, there are;

- League does not affect the market value, probably there is a high dependency between league and the football clubs.
- 19 of 40 football clubs are included in the equation. Barcelona is the most affecting whereas Bayer Leverkusen is the least affecting football club.
- 4 of 6 age categories are included in equation. Being in the 20-21 age is the most effective, while being over 30 has the least contribution.
- Height is always included in the equation. Being between 180 and 184 cm has the most contribution whereas being between 170 and 174 has the least contribution.
- All dominant foot categories are included in the equation. Being right footed has more contribution than being left footed.
- 15 of 28 nationalities are included in the equation. Being an English footballer has the most contribution whereas being a Swiss footballer has the least contribution.
- Outfitters are not included in equation, probably due most of them were Nike or Adidas.
- All played match quantity categories are included in the equation.
- Contribution to goal affects the market value
- Card numbers do not affect the market value

## 4- CONCLUSION

In this study, we evaluated football players with several criteria in different significance levels within the consideration of multicollinearity and homoscedasticity. As a result, we have achieved to build a regression model in 0.10 significance level with 52 attributes, %20 MAPE and and 0.86 adjusted $R^2$ value. Although the multicollinearity couldn't be eliminated completely, partially successful model is built. One of the reason that we couldn't eliminate multicollinearity completely is the fact that most of the independent variables are connected to each other in real life. Economically prosper clubs gather the successful and valuable players into their clubs, and mostly the young talents are valuable, such as Kylian Mbabpe and Paulo Dybala. Height of 180 and 184 have been proved to be successful, which can be assumed that taller players lack dribbling skills and shorter players lack the control of the ball in the air. Also as another rule of thumb, premier league is accepted as the most challenging league in the Europe, and it is no surprise that the players that are English are more valuable. Only interesting thing in the study is the fact that card numbers didn't affect the market value of the players; which can be caused by the reason that valuable players act more cautiously to do not get any penalty. Overall, the study could be improved by a more reasonable collection of data.